\documentclass[usenatbib,letters]{mn2e}
\usepackage{graphicx}
\usepackage[usenames]{color}
\usepackage{euscript}
\usepackage{amssymb}
\usepackage{journals}

\definecolor{grey}{rgb}{0.5,0.6,0.7}

\title[The 2nd cE galaxy in the NGC~5846 group]{SDSS~J150634.27+013331.6:
the second compact elliptical galaxy in the NGC~5846 group\thanks{Based on
observations obtained at the German-Spanish Astronomical Center,Calar Alto,
operated by the Max-Planck-Institut fur Astronomie Heidelberg jointly with
the Spanish National Commission for Astronomy.}}
\author[I. V. Chilingarian \& G. Bergond]{Igor V. Chilingarian$^{1,2}$\thanks{E-mail:
    Igor.Chilingarian@astro.unistra.fr, chil@sai.msu.ru} and Gilles Bergond$^{3,4}$\\
$^{1}$Centre de donn\'ees astronomiques de Strasbourg, Observatoire
astronomique de Strasbourg,
UMR~7550, \\\ \ Universit\'e de Strasbourg / CNRS, 11 rue de l'Universit\'e, 67000 Strasbourg, France\\
$^{2}$Sternberg Astronomical Institute, Moscow State University, 13 Universitetskij prospect, 119992, Moscow, Russia\\
$^{3}$Calar-Alto Observatory, Centro Astron\'omico Hispano Alem\'an, C/
Jes\'us Durb\'an Rem\'on, 2-2 04004 Almeria, Espa\~na \\
$^{4}$Instituto de Astrof\'isica de Andaluc\'ia (CSIC), Glorieta de la
Astronom\'\i a s/n, 18008 Granada, Espa\~na
}
\begin{document}

\date{Accepted 2010 Mar xx. Received 2010 Mar 8; in original form 2010
Feb 5}

\pagerange{\pageref{firstpage}--\pageref{lastpage}} \pubyear{2009}

\maketitle

\label{firstpage}

\begin{abstract} 
We report the discovery of the second compact elliptical (cE) galaxy
SDSS~$J150634.27+013331.6$ in the nearby NGC~5846 group by the Virtual
Observatory (VO) workflow . This object ($M_B = -15.98$~mag, $R_e =
0.24$~kpc) becomes the fifth cE where the spatially-resolved kinematics and
stellar populations can be obtained. We used archival HST WFPC2 images to
demonstrate that its light profile has a two-component structure, and
integrated photometry from GALEX, SDSS, UKIDSS, and Spitzer to build the
multi-wavelength SED to constraint the star formation history (SFH). We observed
this galaxy with the PMAS IFU spectrograph at the Calar-Alto 3.5m telescope
and obtained two-dimensional maps of its kinematics and stellar population
properties using the full-spectral fitting technique. Its structural,
dynamical and stellar population properties suggest that it had a
massive progenitor heavily tidally stripped by NGC~5846. 
\end{abstract}


\section{Introduction} 

In the magnitude-mean surface brightness diagram and the Fundamental Plane
\citep[FP,][]{DD87}, dwarf and giant early-type galaxies seem to form two
distinct sequences joining at around $M_B = -18$~mag \citep[see][and
references therein]{KFCB09}. However, this bi-modal distribution can be
explained as a projection of the two known monotonous relations of other
structural properties of early-type galaxies as functions of a galaxy
luminosity on-to this parameter space: (a) light profile concentration index
and (b) central surface brightness \citep{GG03,HMI03,KDG03,Ferrarese+06}.
Only objects classified as compact elliptical (cE) or ultra-compact dwarf
(UCD, \citealp{MHI02,Drinkwater+03}) galaxies strongly depart from these 
relations.

They represent the two classes of galaxies supposedly forming by tidal
threshing of more massive progenitors \citep{BCDG01,BCDS03}, i.e. they must
have sharply decreased their stellar masses during the evolution.
Both cE and UCD
classes are represented by only a few dozens of known members including
several transitional cE/UCD objects discovered recently
\citep{CM08,Price+09}. Since all these objects are very dense and small,
much higher stellar velocity dispersions are required to keep them in
equilibrium compared to dwarf elliptical (dE) or dwarf spheroidal (dSph)
galaxies of similar luminosities, thus putting them above the locus of dEs
on the $\sigma$ vs $M_B$ \citep{FJ76} relation. Stellar population
properties of cEs and UCDs are very different from typical dE/dSph
usually being very old (with rare exceptions such as Messier~32) and notably
more metal-rich.

Among known compact elliptical galaxies only M~32 (Local group),
NGC~4486B (Virgo cluster), NGC~5846A (NGC~5846 group), and possibly
ACO~3526~$J124853.91-411905.8$ (Centaurus cluster) reside sufficiently
nearby to allow spatially-resolved studies of their kinematics and stellar
populations using ground-based telescopes. They were considered unique
objects until the recent discovery
\citep{Mieske+05,Chilingarian+07,Price+09,Chilingarian+09} of cEs located at
a distance of the Coma cluster or further, which are, however, spatially unresolved
for ground-based optical observations.

In this \textit{Letter} we report the detection of the fifth\footnote{The
preliminary data analysis \citep{SmithCastelli+10} for the two candidate cEs
in the nearby Antlia cluster \citep{SCFRB08} at $d \approx 35$~Mpc
confirms their membership in the cluster thus extending the sample of nearby
cEs to seven objects.} nearby cE
galaxy made by the Virtual Observatory (VO) fed workflow, which became the
second object of this class in the NGC~5846 group. We study its internal
properties using 3D-spectroscopy and datasets at different wavelength
domains available in the VO and data archives.

\section{Data and Techniques Used}

\citet{Chilingarian+09} describe a VO workflow constructed to search cE
galaxies in nearby clusters. We extended it in order to detect cE candidates
also in nearby groups, which would have higher extent on the sky because of
smaller distances and, therefore, require different settings of the {\sc
SExtractor} software \citep{BA96}. To test the modified workflow, we decided
to use HST images of the central part of the Virgo cluster and the NGC~5846
group known to contain ``legacy'' cEs, NGC~4486B and NGC~5846A.
Surprisingly, the workflow detected a new compact object in the HST WFPC2 images
of the NGC~5846 group 3.1~arcmin south-east of the group centre, which
turned to have a spectrum in the Sloan Digital Sky Survey Data Release 7
(SDSS DR7, \citealp{SDSS_DR7}), proving its membership in the group. The
galaxy is identified as SDSS~J150634.27+013331.6, we will call it NGC~5846cE
throughout the rest of the \textit{Letter}. Recently, NGC~5846cE was
mentioned by \citet{EZ10} where it was classified as an UCD.

The NGC~5846 group, the third massive structure in the local Universe after
the Virgo and Fornax clusters, has been intensively studied in the past and,
therefore, numerous complementary datasets in different wavelength domains
are available in the VO. The group is located at a distance of
26.1~Mpc in the Virgo {\sc III} cloud of galaxies \citep{Tully82,EZ10}
corresponding to a spatial scale 126~pc~arcsec$^{-1}$ and a distance modulus
$32.08$~mag.

\subsection{Photometric Data}

We used the calibrated optical WFPC2 HST images in $F555W$ and $F814W$
(total integration times 2200 and 2300 sec) available from the Hubble Legacy
Archive\footnote{http://hla.stsci.edu/} and found by the cE search workflow
to studying the internal structure of NGC~5846cE. The galaxy has a small
size on the sky, therefore we used other data sources only for the
integrated photometric measurements. All photometric data provided in this
\textit{Letter} are corrected for the Galactic extinction \citep{SFD98}.

The NGC~5846 group is included in the footprints of (1) the GR4 Data Release
of the Medium Imaging Survey (MIS) by the Galaxy Evolution Explorer (GALEX)
and (2) the Data Release 6plus (DR6+) of the Large Area Survey (LAS) of the
UKIRT Infrared Deep Sky Survey (UKIDSS, \citealp{Lawrence+07}), thus providing
photometric measurements in far-UV, near-UV, and four near-IR broadband
filters $YJHK$ in addition to the 5-band optical $ugriz$ photometry from
SDSS DR7. We took Petrosian magnitudes from SDSS and UKIDSS, applying
Vega-to-$AB$ zero-point correction for the latter ones according to
\citet{HWLH06}, and total FUV and NUV magnitudes from GALEX.

There are publicly available archival Spitzer Space Telescope images
obtained with the Infrared Array Camera (IRAC) in four photometric bands
centered at 3.6, 4.5, 5.8, and 8.0~$\mu$m. We obtained total $AB$ magnitudes of
NGC~5846cE in the IRAC bands using {\sc SExtractor} (\textit{MAG\_AUTO} and
\textit{MAGERR\_AUTO} parameters).

Several central pixels of the galaxy image in all HST WFPC2 frames are
saturated, therefore no analysis of the inner region is possible.
The images were background-subtracted using {\sc SExtractor}. Then we
obtained light profiles of NGC~5846cE in both photometric bands by fitting
elliptical isophotes with free orientation, ellipticity, and disky/boxy
parameters using the {\sc stsdas.analysis.isophote.ellipse} task in the {\sc
iraf} data processing environment.

\begin{figure}
\includegraphics[width=0.95\hsize]{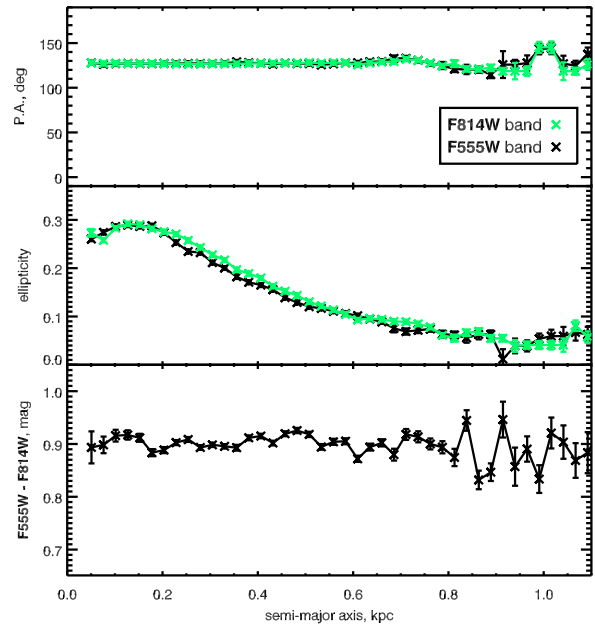}
\caption{Radial behaviour of the major axis positional angle (top),
ellipticity (middle) of the isophotes of NGC~5846cE from the
HST WFPC2 $F555W$ and $F814W$-band images shown in black and green respectively. 
The bottom panel displays the reconstructed $F555W - F814W$ colour 
profile.\label{figpaell}}
\end{figure}

In Fig~\ref{figpaell} we present the radial behaviour of ellipticity $e = 1
- b/a$, and positional angle (top and middle panels) and the $F555W - F814W$
colour profile. The positional angle remains stable at $PA = 127$~deg at all
radii. The galaxy has very round outer isophotes ($e \sim 0.05$) becoming
significantly prolate inwards with the ellipticity reaching ($e = 0.3$) at
$r = 0.125$~kpc$ = 1$~arcsec. Closer to the centre the ellipticity starts to
decrease, however, we could not measure it at $r<0.3$~arcsec due to the
saturation mentioned above. The radial behaviour of $PA$ and $e$ is
identical in the two photometric bands. The isophotes remain purely
elliptical without any signature of diskyness/boxiness. 

The reconstructed colour profile is completely flat having a value of
$F555W-F814W=0.90$~mag. We computed a 2-dimensional colour map of NGC~5846cE
applying the Voronoi adaptive binning \citep{CC03} with a target
signal-to-noise ratio 100 in $F555W$. It contains no statistically
significant deviations from the same constant level. The unsharp masking
technique using the elliptical Gaussian smoothing kernel with the parameters
corresponding to the inner isophotes did not reveal any embedded structures
in NGC~5846cE.

\begin{figure}
\includegraphics[width=\hsize]{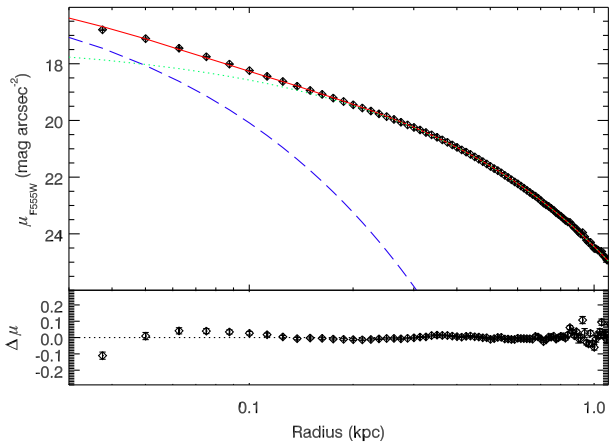}
\caption{The two-component NGC~5846cE light profile decomposition. The top
panel displays the brightness profile shown with black diamonds, and the
two components represented by the blue dashed and green dotted lines
for inner and outer S\'ersic profiles correspondingly. 
The bottom panel shows the fitting residuals.
\label{figdecompos}}
\end{figure}

The light profile in the outer part ($r > 2.0$~arcsec) is perfectly
approximated by the \citet{Sersic68} profile having $n\approx1.5$, but in
this case there is a light excess in the inner part of the galaxy.
Therefore, we performed the structural decomposition of the light profile
using two S\'ersic components and fitting their parameters non-linearly. We
used a modified version of the algorithm presented in
\citet{Chilingarian+09b}. The obtained structural parameters of NGC~5846cE
in the $F555W$ band are presented in Table~\ref{tabdecomp}. Parameters of
the inner component are badly constrained because of the saturated nucleus.
The light profile and its components are shown in Fig~\ref{figdecompos}, the
scale is set logarithmic on the $r$ axis in order to emphasize the inner
component which becomes barely visible in linear scale. The total
reconstructed absolute magnitude of NGC~5846cE is $M_{F555W}=-17.07 \pm 0.10$~mag. In
order to transform these measurements into the Johnson $B$ band, the $B -
F555W = 0.92$~mag transformation should be applied according to
\citep{JGA06} and $g - r = 0.78$~mag from SDSS DR7. The absolute $B$ 
band magnitude computed from the SDSS DR7 photometry is $M_B = -15.98$~mag.

\begin{table}
\caption{NGC~5846cE $F555W$-band light profile decomposition using the 2-component
model. }
\begin{tabular}{lcc}
\hline
\hline
& S\'ersic$_{\rm{in}}$ & S\'ersic$_{\rm{out}}$ \\
\hline 
$r_{e}$ kpc & 0.038 $\pm$ 0.012 & 0.291 $\pm$ 0.009 \\
$r_{e}$ arcsec & 0.30 $\pm$ 0.10 & 2.33 $\pm$ 0.07 \\
$n$  & 1.56 $\pm$ 0.26 & 1.41 $\pm$ 0.02\\
$\mu_{e}$ mag~arcsec$^{-2}$ & 17.48 $\pm$ 1.05 & 19.86 $\pm$ 0.13\\
$\langle\mu\rangle_{e}$ mag~arcsec$^{-2}$ &  16.56 $\pm$ 1.05 & 19.00 $\pm$ 0.13 \\
$M_{F555W}$~mag & $-14.92$ & $-16.91$ \\
\hline
\hline
\end{tabular}
\label{tabdecomp}
\end{table}

\subsection{Spectroscopic Data}

We observed NGC~5846cE with the Potsdam Multi-Aperture Spectrograph (PMAS,
\citealp{Roth+05}) mounted at the 3.5-m telescope of the Calar-Alto
Observatory on 2009-Apr-22 in the framework of our observing programme
``3D-spectroscopy of dE galaxies with kinematically-decoupled cores'' (P.I.:
GB). NGC~5846cE was an extra target observed in the morning. We made three
30-min long exposures in the LArr mode with the Integral-Field Unit (IFU)
having 16$\times$16 square 1$\times$1~arcsec lenses. The FWHM seeing was
about $1.8$~arcsec. We used the V1200 grism providing the resolving power
$R\approx2800$ in the wavelength range 4720--5420\AA.

For the data reduction, we used the generic IFU data reduction pipeline
implemented in {\sc idl}. A brief description of the data reduction
steps can be found in \cite{CPSA07}. We performed the flux calibration using
the observations of the $HZ~44$ spectrophotometric standard star. The night
sky spectrum was reconstructed from the lenses located in the outer parts of
the field of view ($r > 6$~arcsec). For the analysis we used only the inner
8$\times$8~arcsec fragment of the field of view centered on NGC~5846cE. We
applied the Voronoi adaptive 2D binning to the final sky-subtracted and
flux-calibrated data cube.

\begin{figure}
\includegraphics[width=0.49\hsize]{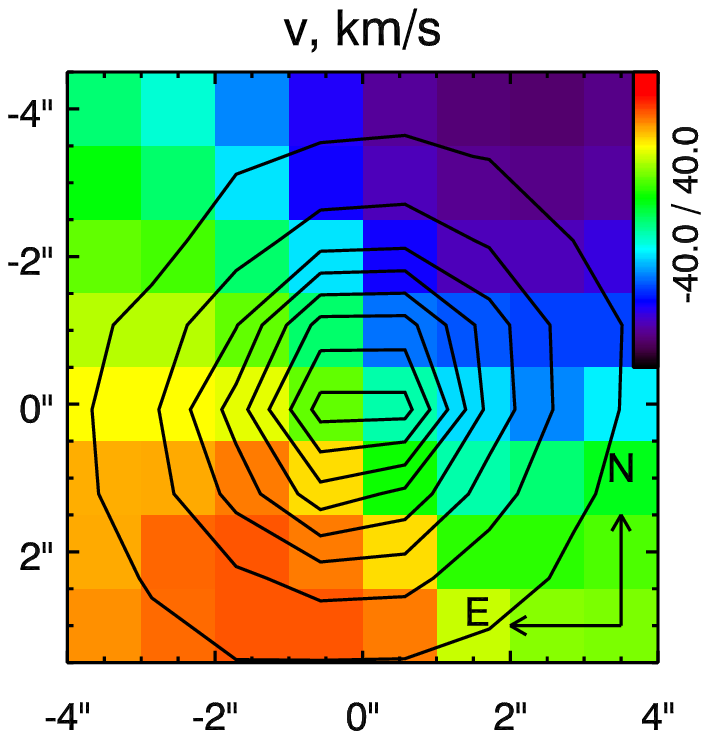}
\includegraphics[width=0.49\hsize]{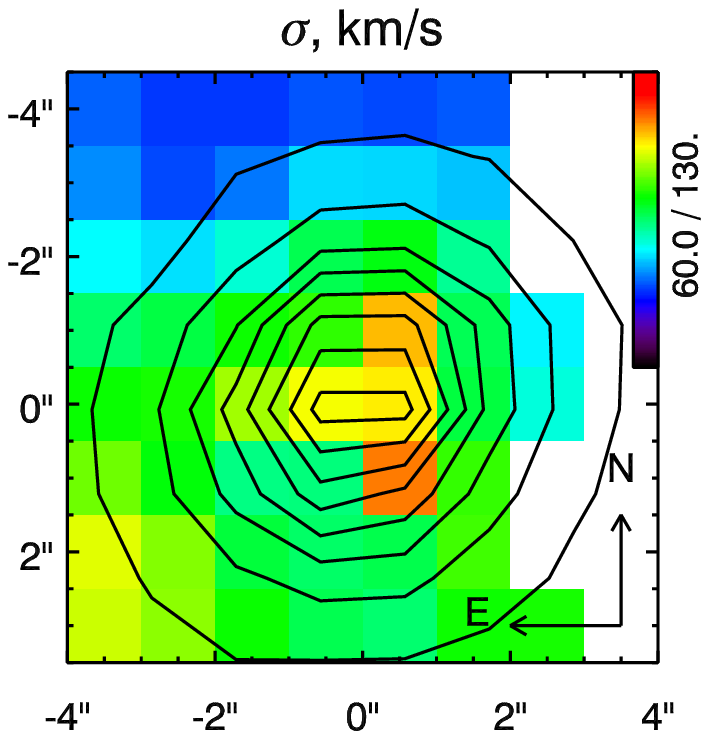}\\
\includegraphics[width=0.49\hsize]{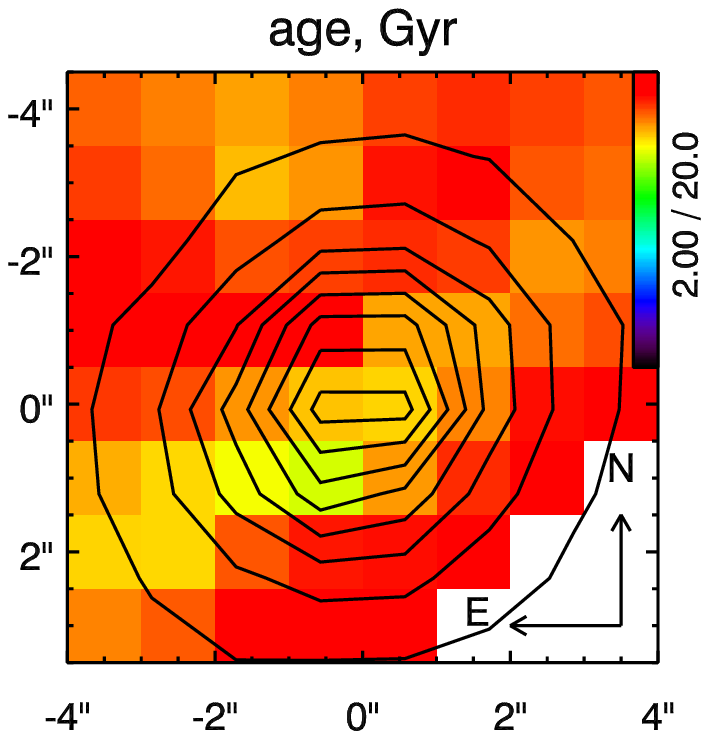}
\includegraphics[width=0.49\hsize]{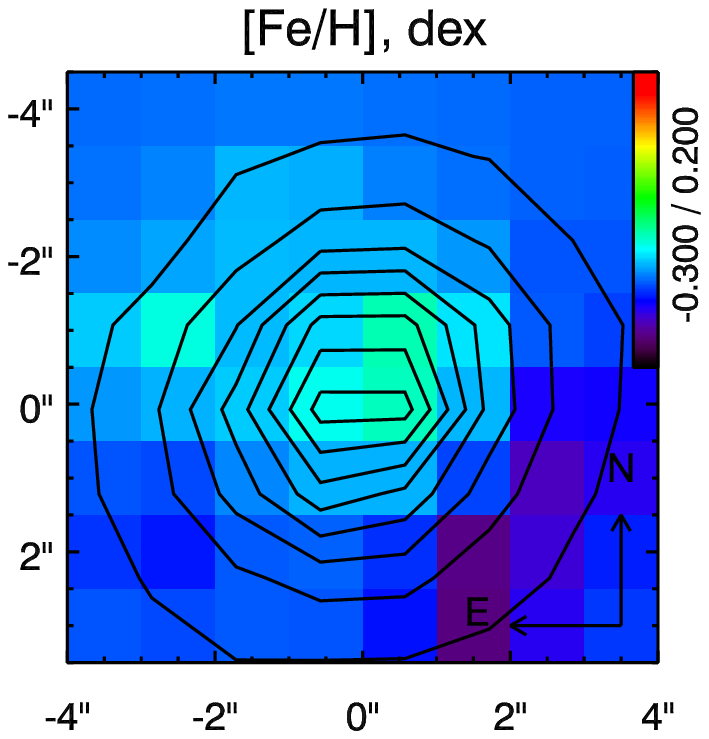}\\
\caption{Two-dimensional maps of kinematics and stellar populations of
NGC~5846cE obtained from the full-spectral fitting of the PMAS IFU dataset:
radial velocity and velocity dispersion (upper row), SSP-equivalent age and
metallicity (lower row).
\label{figifumaps}}
\end{figure}

We exploited the {\sc NBursts} full spectral fitting technique \citep{CPSK07}
to fit the binned PMAS dataset, and derived maps of stellar radial velocity,
velocity dispersion, age and metallicity of stellar populations presented in
Fig~\ref{figifumaps}. We used the grid of simple stellar populations (SSP)
computed with the {\sc pegase.hr} code \citep{LeBorgne+04}, therefore
the derived stellar population parameters are SSP-equivalent. The target
signal-to-noise ratios of 10 and 15 per pixel were used for computing
kinematical and stellar population maps respectively.

Despite its almost round outer isophotes, the galaxy exhibits a regular
velocity field with a significant major-axis rotation ($v_{r} \sin i =
$40~km~s$^{-1}$). We are probably reaching the maximum of rotation already
at $r = 3\dots4$~arcsec comparable to M~32 \citep[e.g.][]{SP02},
however, deeper observations with higher spatial resolution are required to
give decisive conclusions about this point.

The velocity dispersion distribution has a pronounced bump in the centre reaching
$\sigma_{0} = 118$~km~s$^{-1}$ with the values smoothly decreasing outwards down to
75--85~km~s$^{-1}$ at $r=3$~arcsec. Due to the limited spatial resolution of
our data, the real value of the central velocity dispersion is probably
underestimated and it can be much higher as in other nearby cEs
\citep{DBM08}.

Stellar population of the galaxy is very old ($t = 15 \pm 4$~Gyr) and
metal-rich ($[\rm{Z/H}] = -0.04 \pm 0.06$~dex). The metallicity distribution
slightly decreases outwards, however the change is only $-$0.1~dex. The age
map does not exhibit any statistically significant details.

We also used the flux-calibrated SDSS DR7 spectrum of NGC~5846cE ($R=1800$)
obtained in a 3~arcsec-wide circular aperture. We applied the {\sc NBursts}
technique to this spectrum in the wavelength range $3900 - 6800$~\AA\ in order
to perform the cross-check of the kinematical and stellar population
parameters in the central region of the galaxy. The values ($v_r = 1537 \pm
2$~km~s$^{-1}$, $\sigma_0 = 119 \pm 2$~km~s$^{-1}$, $t = 15.3 \pm 0.6$~Gyr,
$[\mathrm{Z/H}] = -0.04 \pm 0.02$~dex) agree remarkably well with those
obtained from the PMAS data. No emission lines are detected ruling out the
ongoing star formation (SF) in the galaxy as well as the presence of ionized gas
in any form.

We also used the measurements of Lick absorption line strength indices
\citep{WFGB94} for Mg$b$, Fe$5270$ and Fe$5335$ provided by the SDSS and
models of \citet{TMB03} in order to estimate the level of $\alpha/$Fe
enhancement of stellar population. We derived $[$Mg/Fe$] = 0.34 \pm
0.05$~dex.

\section{Discussion}

\begin{figure}
\includegraphics[width=\hsize]{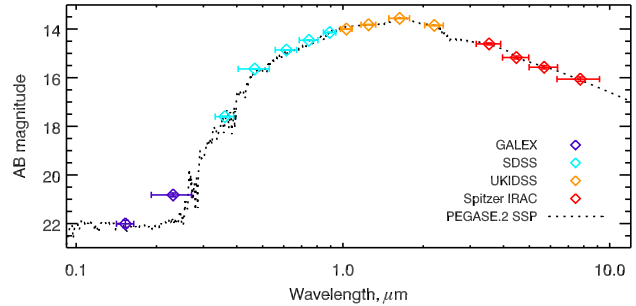}
\caption{Multi-wavelength SED of
NGC~5846cE constructed from the integrated photometry found in the VO,
and the Spitzer Space Telescope archive. Horizontal bars denote
the FWHM of corresponding photometric bands. Dotted black line is a {\sc
PEGASE.2} SSP model without internal extinction corresponding to the
$t = 15$~Gyr and $[$Fe/H$]$=0.0~dex.\label{figsed}}
\end{figure}

In Fig~\ref{figsed} we demonstrate the multi-wavelength spectral energy
distribution (SED) of NGC~5846cE constructed form the integrated photometric
data described in the previous section. We overplot the low-resolution SSP
model computed with {\sc pegase.2} \citep{FR97} for the age and metallicity
of stars close to those obtained from the full spectral fitting of the SDSS
DR7 spectrum of NGC~5846cE. No internal extinction is assumed. One can see a
perfect agreement between the SSP model and the observed galaxy SED at all
wavelengths from 1500\AA\ to 8~$\mu$m. The absence of the UV excess clearly
rules out any possibility of having extended SFH and residual recent SF, the
agreement of all SED details at such a large wavelength basis proves the
lack of dust. All this, in addition to high value of $[$Mg/Fe$]$ abundance
ratios suggest \citep{Matteucci94} that the star formation history of
NGC~5846cE contained the only one very short and intense starburst in the
remote past.

The structural properties and low luminosity place NGC~5846cE between
M~32 and A496cE \citep{Chilingarian+07} on the FP and
its projections. At the same time, its stellar population properties, in
particular very Mg/Fe overabundance, make it resembling extreme cases
like A496g1 and NGC~4486B. Indeed, NGC~5846cE exhibits a two-component light
profile. However, the important difference between this object and most
cE \citep{Graham02,Chilingarian+07} and transitional cE/UCD
galaxies \citep{CM08,Price+09} is that even its outer component is 
compact and bright.

\citet{Chilingarian+09} demonstrate using numerical simulations that the
tidal stripping of a disc (possibly barred) galaxy by the cluster cD
potential is the most plausible formation scenario of cE galaxies. In case
of NGC~5846cE, the role of a cD is played by NGC~5846, a very massive
non-rotating \citep{Emsellem+04} elliptical galaxy. The isophote ellipticity
increasing toward the centre, important major axis rotation, and at the same
time, very regular elliptical isophotes and the absence of embedded
structures, support the scenario of a tidally stripped bar suggested by
simulations by \citep{Chilingarian+09} and presented in their fig.~3. From
the luminosity--metallicity relation presented in the lower panel of fig.~1
from the same paper, we can estimate the luminosity of the NGC~5846cE's
progenitor to be about $M_B = -19$~mag, hence the mass loss due to the tidal
stripping should be a factor of 15.

We can compare the present day stellar mass of NGC~5846cE computed in case
of two different stellar initial mass functions (IMF) similarly to what we
did for UCDs \citep{CCB08}. We use the total
luminosity of the galaxy from SDSS DR7, age and metallicity of its stellar
population from the full spectral fitting and corresponding mass-to-light
ratios provided by {\sc pegase.2}. The results are different by a factor of
almost two: $(4.2 \pm 0.4)\times10^9 M_{\odot}$ and $(2.2 \pm 0.3)\times10^9
M_{\odot}$ for the \citet{Salpeter55} and \citet{Kroupa01} IMFs
correspondingly\footnote{Stellar population parameters remain virtually the
same for the SSP models computed using two different IMFs. All values
provided in the previous section are computed for the Salpeter IMF based SSP
models.}. At the same time, using a simple virial mass estimate $10 R_e
\sigma_v^2 / G$ \citep{Spitzer69} which will probably underestimate the
mass, we get $\sim (5.5 \pm 0.6)\times10^9 M_{\odot}$ for the dynamical mass
assuming the global velocity dispersion $\sigma_v =
100\pm5$~km~s$^{-1}$. Hence, our galaxy may contain about 60~per~cent of
dark matter if we adopt the Kroupa IMF, whereas it becomes almost dark
matter free in case of the Salpeter IMF, making it similar in this respect to
UCDs which are also believed to be tidally stripped objects.

Hence, all observed properties of NGC~5846cE suggest that it was probably
formed by tidal stripping of a massive disky (probably barred) galaxy by
NGC~5846 similarly to compact elliptical galaxies in nearby clusters. Unlike
M~32, NGC~5846cE is an excellent example of a cE very similar to those
observed in galaxy clusters, however, located relatively nearby. It will
thus hopefully allow us to study in detail its internal dynamics and mass
distribution using available observational facilities in order to better
understand tidal stripping of galaxies.

\section*{Acknowledgments}

In this study, we used the UKIDSS DR6plus survey catalogues available
through the WFCAM science archive, SDSS DR7, and GALEX GR4 data. Funding for
the SDSS and SDSS-II has been provided by the Alfred P. Sloan Foundation,
the Participating Institutions, the National Science Foundation, the U.S.
Department of Energy, NASA, the Japanese Monbukagakusho, the Max Planck
Society, and the Higher Education Funding Council for England. The SDSS Web
Site is http://www.sdss.org/. We acknowledge the usage of the {\sc topcat}
software by M.Taylor. This work is based in part on observations made with
the Spitzer Space Telescope, which is operated by the JPL, California
Institute of Technology under a contract with NASA

\bibliographystyle{mn2e}
\bibliography{ngc5846cE}

\label{lastpage}

\end{document}